\pdfoutput=1

\documentclass[11pt, dvipsnames]{article}

\usepackage[final]{acl}

\usepackage{times}
\usepackage{latexsym}

\usepackage[T1]{fontenc}

\usepackage{amsmath}
\usepackage{adjustbox}
\usepackage{amssymb}

\AtBeginDocument{%
  }
\usepackage{graphicx}
\usepackage{subfig}

\usepackage[utf8]{inputenc}

\usepackage{microtype}

\usepackage{inconsolata}

\usepackage{graphicx}
\usepackage{algorithm}
\usepackage{algpseudocode}
\usepackage{amsmath}
\usepackage{dblfloatfix} 
\usepackage{multirow}
\usepackage{booktabs}
\usepackage{verbatim}
\usepackage{balance}
\usepackage{array}
\usepackage{float}
\usepackage{adjustbox}
\usepackage{afterpage}
\usepackage[table]{xcolor}
\algdef{SE}[VARIABLES]{Variables}{EndVariables}
   {\algorithmicvariables}
   {\algorithmicend\ \algorithmicvariables}
\algnewcommand{\algorithmicvariables}{\textbf{Declare}}

\PassOptionsToPackage{table}{xcolor}

\usepackage{times}
\usepackage{latexsym}
\usepackage[T1]{fontenc}
\usepackage[utf8]{inputenc}
\usepackage{microtype}
\usepackage{inconsolata}
\usepackage{graphicx}
\usepackage[table]{xcolor}
\usepackage{amsmath}
\usepackage{amssymb}

\usepackage{wrapfig}

\title{Long Context Modeling with Ranked Memory-Augmented Retrieval}

\author{
  Ghadir Alselwi\textsuperscript{1},
  Hao Xue\textsuperscript{1,2,3},
  Shoaib Jameel\textsuperscript{4},
  Basem Suleiman\textsuperscript{1}, \\
  \textbf{Flora D. Salim}\textsuperscript{1},
  \textbf{Imran Razzak}\textsuperscript{5,1} \\
  \textsuperscript{1}University of New South Wales, Sydney, NSW, Australia \\
  \textsuperscript{2}Hong Kong University of Science and Technology (Guangzhou) \\
  \textsuperscript{3}The Hong Kong University of Science and Technology \\
  \textsuperscript{4}University of Southampton, Southampton, UK \\
  \textsuperscript{5}Mohamed Bin Zayed University of Artificial Intelligence, Abu Dhabi, UAE \\
  \texttt{\{g.alselwi,b.suleiman,flora.salim\}@unsw.edu.au} \\
  \texttt{haoxue@hkust-gz.edu.cn, M.S.Jameel@southampton.ac.uk, imran.razzak@mbzuai.ac.ae} \\
}

\begin{document}
\maketitle

\begin{abstract}
Effective long-term memory management is crucial for language models handling extended contexts. We introduce the Enhanced Ranked Memory Augmented Retrieval (\textbf{ERMAR}) framework, which dynamically ranks memory entries based on relevance. Unlike prior models, ERMAR employs a novel relevance scoring mechanism and a pointwise re-ranking model for key-value embeddings, inspired by learning-to-rank techniques in information retrieval. By integrating historical usage patterns and adaptive retrieval, ERMAR achieves state-of-the-art results on standard benchmarks, demonstrating superior scalability and performance in long-context tasks. Code will be released at \url{https://github.com/cruiseresearchgroup/ERMAR}.
\end{abstract}

\section{Introduction}
\begin{figure*}
    \centering
    \includegraphics[scale=0.6]{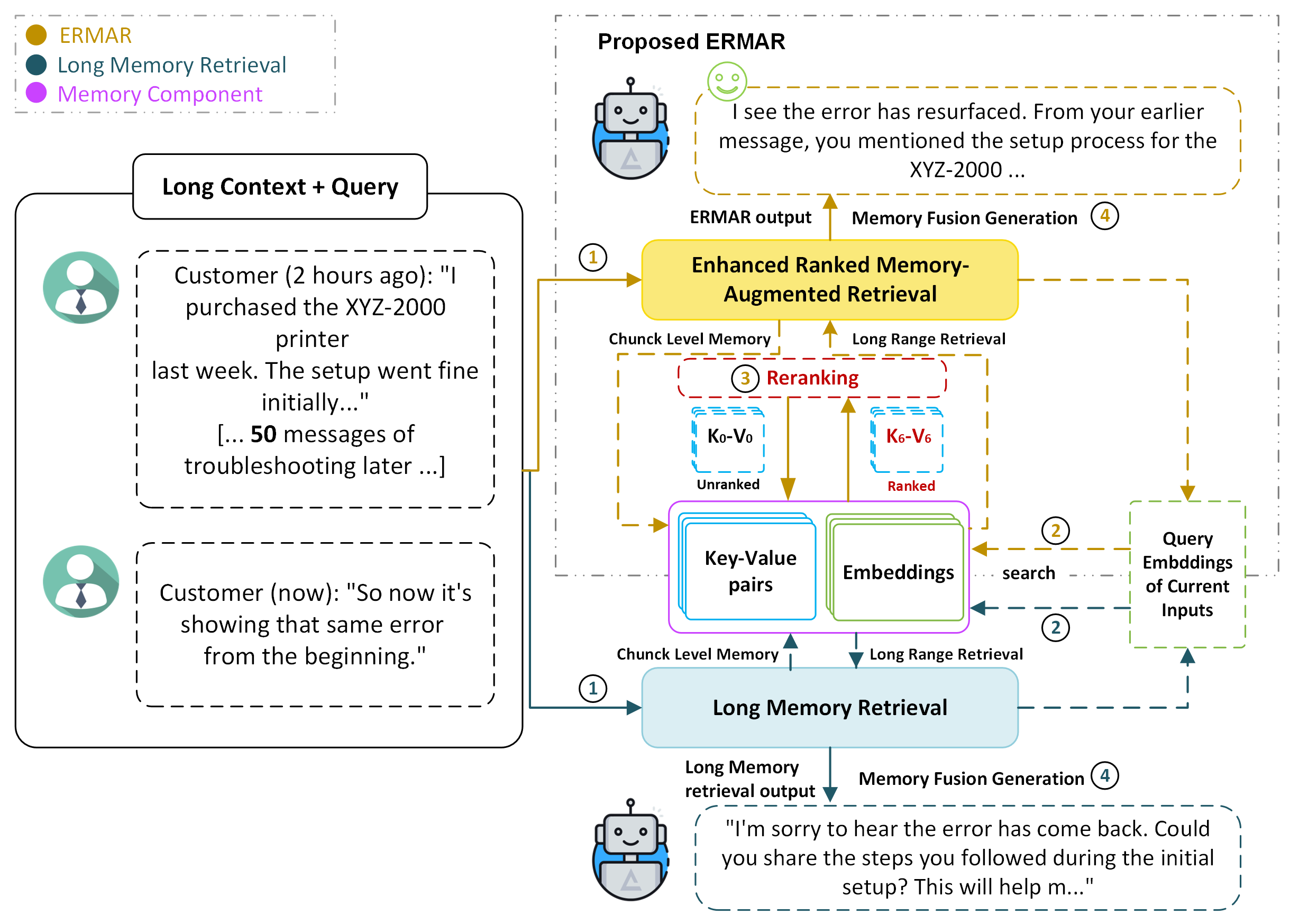}
    \caption{Motivating example of the ERMAR retrieval workflow.  \textit{Left}: A long customer support conversation where the current query  references information from much earlier context.  \textit{Right}: The four-stage workflow --- \textcircled{1}~Long Memory Retrieval,  \textcircled{2}~Search, \textcircled{3}~Reranking, \textcircled{4}~Memory Fusion  Generation. The ERMAR output (top) correctly references earlier context, whereas  the output without reranking (bottom) produces a generic response.}
    \label{fig:toyexample}
\end{figure*}

Large Language Models (LLMs) face a fundamental limitation in processing long-context scenarios due to the quadratic complexity of attention mechanisms and increasing memory demands during generation~\cite{vaswani2017attention, tworkowski2024focused}. Consider a scenario in an automated customer service system: \textit{A customer reports an issue with their printer, referencing a setup process from a previous conversation that occurred two hours ago. After 50 messages of troubleshooting, the customer mentions that the same error from the beginning has resurfaced. Traditional LLMs, constrained by their context window, would struggle to access the crucial earlier context about the initial setup process, leading to inconsistent or incomplete responses, Figure~\ref{fig:toyexample}.} It is well known that handling extended contexts remains a significant challenge, particularly in applications requiring document analysis and sustained dialogue interactions.

The recent MemLong~\cite{liu2024memlong} architecture stores and accesses historical context through basic chunk-level memory operations. The memory bank model is a large, non-trainable store of past context representations. Instead of re-computing representations for all past tokens every time, these representations are pre-computed and stored. Given the current context, MemLong retrieves relevant segments from the memory bank. It uses a dot product similarity search to find the memory entries most related to the current context. This allows the model to focus only on the most pertinent past information, rather than processing the entire history. However, its treatment of all key-value (K-V) pairs with equal weight, regardless of their contextual relevance, often leads to information overload and reduced retrieval precision. This limitation becomes particularly evident in scenarios requiring context management.

We have developed a novel model that addresses the aforementioned limitations by building upon Memlong~\cite{liu2024memlong}, a publicly available baseline on GitHub\footnote{https://github.com/Bui1dMySea/MemLong}. Our \textbf{Enhanced Ranked Memory Augmented Retrieval} (\textbf{ERMAR}) model has a novel relevance scoring mechanism that fundamentally improves context retrieval and utilization for K-V embeddings. Unlike MemLong, ERMAR employs multiplication \cite{cao2007learning} to compute relevance scores, enabling a more nuanced and context-aware assessment of semantic alignment between queries and stored memory. ERMAR also incorporates a re-ranking mechanism that dynamically reorders K-V embeddings based on their relevance scores, ensuring that the most pertinent information is prioritized during retrieval. This re-ranking process, combined with an adaptive retrieval system that integrates historical usage patterns, allows ERMAR to capture subtle contextual relationships better and refine memory prioritization. As shown in Figure~\ref{fig:toyexample}, ERMAR processes incoming queries and long-context conversations through a novel ranking architecture, employing K-V pairs ranking (\scalebox{0.9}{$K_0$-$V_0$, $K_1$-$V_1$,...,$K_i$-$V_i$}) and corresponding embeddings to perform semantic search and ranking of relevant historical information.

Our novel ERMAR model introduces three key improvements: (i) A semantic similarity metric to measure contextual alignment between query embeddings and key-value pairs; (ii) A weighted scoring function that considers content similarity and contextual relevance; and (iii) Integration of historical usage patterns to refine relevance assessment.

\section{Related Work}
The challenge of enabling language models to effectively process extended contexts has driven research across multiple domains. We organize related work into key areas that inform our ERMAR framework.

\subsection{Memory-Augmented Neural Networks}

Early architectures like Neural Turing Machines~\cite{graves2014neural} and Differentiable Neural Computers~\cite{graves2016hybrid} introduced external memory beyond model parameters. Recent models for long-context language modeling include RetroMAE~\cite{xiao2022retromae}, which struggled with semantic coherence, and Memorizing Transformers~\cite{wu2022memorizing}, which improved cross-sequence information storage but faced scalability limits. MemLong~\cite{liu2024memlong} stores historical context via chunk-level memory operations, but its uniform treatment of key-value pairs can cause information overload and reduced retrieval precision.

\subsection{Retrieval-Augmented Generation and Long-Context Modeling}
Retrieval-augmented generation (RAG)\cite{lewis2020retrieval} pioneered integrating dense passage retrieval with generative models, while Fusion-in-Decoder (FiD)\cite{izacard2020leveraging} improved efficiency through independent passage processing. REALM~\cite{guu2020retrieval} introduced end-to-end learning of retrieval and generation, and DPR~\cite{karpukhin2020dense} established dense passage retrieval standards. However, these methods typically focus on static corpora rather than dynamic context-aware memory management.
The quadratic complexity of attention mechanisms has driven research into efficient long-context architectures. Sparse attention mechanisms such as Longformer~\cite{beltagy2020longformer} and BigBird~\cite{zaheer2020big} reduce computational complexity while maintaining model capabilities through selective attention patterns. Position encoding adaptations like RoPE~\cite{su2024roformer} and ALiBi~\cite{press2022trainshorttestlong} have enhanced models' ability to handle longer sequences. YARN~\cite{peng2023yarn} further advanced this through dynamic position embeddings, demonstrating reliable generalization up to 128k tokens.
\subsection{Learning-to-Rank and Information Retrieval}
Our ERMAR framework draws inspiration from learning-to-rank techniques in information retrieval. Traditional ranking approaches like BM25 struggle with semantic similarity, leading to neural ranking models using learned representations. Pointwise ranking approaches~\cite{cao2007learning} predict relevance scores for query-document pairs, directly inspiring our relevance scoring mechanism. Dense retrieval methods like DPR~\cite{karpukhin2020dense} and ColBERT~\cite{khattab2020colbert} demonstrate the effectiveness of learned dense representations for ranking through similarity scores—a paradigm we adapt for memory entry ranking.
BERT-based re-ranking models~\cite{nogueira2019passage} show that sophisticated re-ranking significantly improves retrieval quality. This two-stage retrieve-then-rerank paradigm directly influences our design, where we first retrieve candidate memory entries and then apply learned re-ranking.

\subsection{Current Limitations and Gaps}
Despite significant progress, current approaches face limitations that motivate ERMAR:

\textbf{Static Memory Management}: Most approaches use fixed memory structures that do not adapt to content importance or usage patterns, leading to inefficient memory utilization.

\textbf{Uniform Memory Treatment}: Existing methods treat all memory entries equally, lacking mechanisms to prioritize more relevant information.

\textbf{Limited Semantic Understanding}: Memory systems often rely on simple similarity metrics without considering contextual nuance or temporal relevance.

\textbf{Scalability Trade-offs}: Current approaches face difficult trade-offs between memory capacity, retrieval accuracy, and computational efficiency.

Our ERMAR framework addresses these limitations through dynamic relevance scoring, adaptive memory management, and sophisticated re-ranking mechanisms inspired by information retrieval techniques, providing a more principled approach to long-context memory management.

\section{Our Novel ERMAR Model}

\begin{figure}[t]
    \centering
    \includegraphics[scale=0.56]{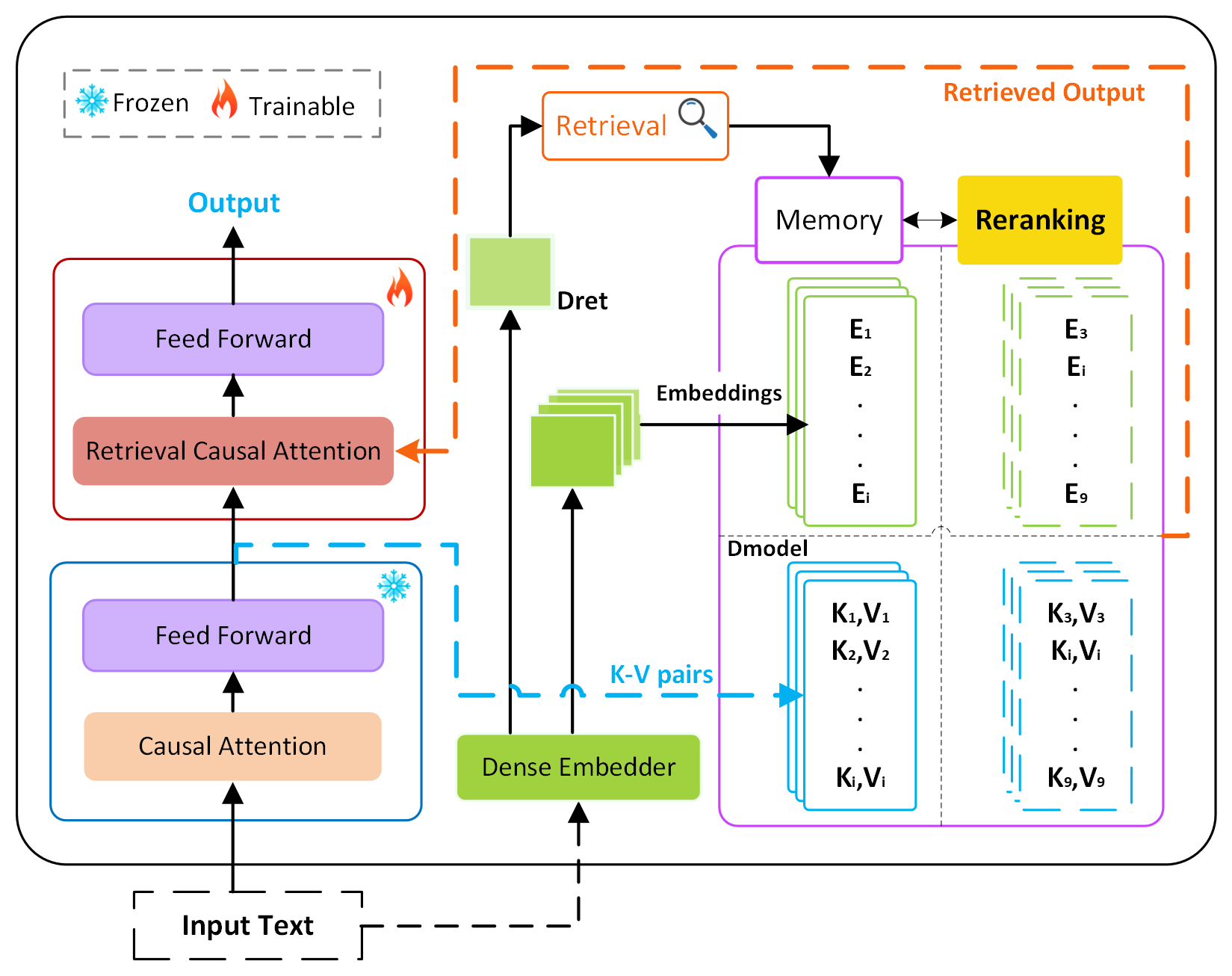}
    \caption{Architecture of the ERMAR model. The frozen lower block (blue dashed, snowflake) produces hidden states  stored as key--value pairs ($d_\text{model}$) in the Memory Bank, while  the \textbf{Dense Embedder} encodes the input into chunk-level retrieval  embeddings ($d_\text{ret}$). The \textbf{Reranking} module scores and ranks all stored entries, and the \textbf{Retrieval} module injects the top-$K$ ranked pairs into the trainable upper block via \textbf{Retrieval Causal Attention} (orange arrow) to condition the final output. Frozen and trainable components are marked by snowflake and flame icons respectively.}
    \label{fig:arch}
\end{figure}

Figure~\ref{fig:toyexample} presents a concrete motivating example of how ERMAR's  contextual ranking mechanism operates in practice, while Figure~\ref{fig:arch}  illustrates the overall architecture of the proposed framework. As shown in Figure~\ref{fig:toyexample}, ERMAR processes a long-context query  through four sequential stages. First, the \textbf{Long Memory Retrieval} module encodes the full input history  into chunk-level key--value (KV) pairs stored in a memory bank, alongside  embeddings that capture semantic representations of each chunk. Second, when a new query arrives, its embedding is used to \textbf{search} the  memory bank for candidate KV pairs via similarity matching. Third, the retrieved candidates pass through the \textbf{Reranking} module --- the  core novelty over MemLong --- which assigns relevance scores to each KV pair and  re-orders them by contextual importance, elevating high-relevance entries  ($K_6$-$V_6$, ranked) above lower-relevance ones ($K_0$-$V_0$, unranked). Finally, only the top-ranked KV pairs are passed to \textbf{Memory Fusion Generation}, which conditions the model's output on the most relevant historical  context.

This design directly addresses the information dilution problem in long contexts:  without reranking (bottom of Figure~\ref{fig:toyexample}), the model produces a  generic response that fails to reference the earlier printer setup context; with  reranking (ERMAR output, top), the model correctly retrieves and incorporates the  relevant earlier information.

As shown in Figure~\ref{fig:arch}, the ERMAR architecture consists of three  interacting components: a \textbf{frozen lower transformer block}, a  \textbf{trainable upper block} with retrieval-augmented attention, and a  \textbf{memory retrieval module} with reranking. Input text is processed by the frozen lower block (Causal Attention and Feed  Forward layers), whose hidden states form the key--value pairs  $\{(K_j, V_j)\} \in \mathbb{R}^{d_\text{model}}$ stored in the Memory Bank. In parallel, the \textbf{Dense Embedder} independently encodes the same input into  chunk-level retrieval embeddings $E_1, \ldots, E_i \in \mathbb{R}^{d_\text{ret}}$,  which serve as the semantic index for similarity-based retrieval. The \textbf{Reranking} module then scores all stored embeddings and KV pairs using  Equation~\ref{eq:relevance_score}, re-ordering them by relevance to the current  query and isolating the top-ranked subset (right stacks in Figure~\ref{fig:arch}). The \textbf{Retrieval} module performs top-$K$ selection over this ranked subset and injects the selected KV pairs into the trainable upper block via  \textbf{Retrieval Causal Attention} (orange arrow), producing a context-enriched representation from which the final output is generated.

ERMAR maintains consistency through frozen lower layers and selective parameter  updates. Specifically, ERMAR (i)~stores important information from earlier parts  of the text, (ii)~assigns relevance scores to stored information based on its  importance to the current context, and (iii)~retrieves only the most relevant  historical information when needed. The relevance scoring mechanism is analogous  to the attention operation, enabling the model to focus on salient parts of the  memory. There is also a ``loose'' pointwise connection: while relevance-based  reranking acts as an auxiliary mechanism that enhances memory retrieval quality,  the primary optimization target remains next-token prediction likelihood.

Let $\mathcal{V}$ be a finite vocabulary, and 
$\mathbf{x} = (x_1, \ldots, x_n) \in \mathcal{V}^n$
be a token sequence with a preceding context $\mathbf{x}_{<i}$. 
The embedding function 
$\mathcal{E}: \mathcal{V}^* \rightarrow \mathbb{R}^{d_\text{ret}}$
maps sequences to a retrieval space. 
We introduce a memory function $\mathcal{M}$ augmented with a relevance scoring mechanism,
\begin{equation}
\small
\scalebox{1.1}{$
\mathcal{M}: 
\underbrace{\mathbb{R}^{d_\text{model}}}_{\text{keys}} 
\times 
\underbrace{\mathbb{R}^{d_\text{model}}}_{\text{values}} 
\times 
\underbrace{\mathbb{R}^{d_\text{ret}}}_{\text{embeddings}} 
\rightarrow 
\mathcal{S}
$},
\end{equation}
where $d_\text{model}$ denotes the transformer hidden dimension, $d_\text{ret}$ the retrieval embedding dimension, and $\mathcal{S}$ the space of scored memory items (see Appendix~\ref{sec:embedding_dim_sensitivity} for embedding dimensionality analysis).

\textbf{Relevance Score.} 
Given a query embedding $\mathbf{q} \in \mathbb{R}^{d_\text{ret}}$ and a matrix of key embeddings 
$\mathbf{K} = [k_1, \ldots, k_m] \in \mathbb{R}^{m \times d_\text{ret}},$
\noindent where each row $k_i \in \mathbb{R}^{d_\text{ret}}$ corresponds to a key embedding, the relevance score is defined as:
\begin{equation}
\small
\alpha(\mathbf{q}, \mathbf{K}) = \text{softmax}\left(\frac{\mathbf{q}\mathbf{K}^\top}{\sqrt{d_\text{ret}}}\right),
\label{eq:relevance_score}
\end{equation}
\noindent where $\sqrt{d_\text{ret}}$ normalizes the similarity scores to prevent excessively large values. 

The resulting vector $\alpha(\mathbf{q}, \mathbf{K})$ represents a probability distribution over the keys, where each entry $\alpha_i$ denotes the relative importance (attention weight) of the $i$-th key with respect to the query $\mathbf{q}$. 
The final context or weighted output is obtained as
$\mathbf{o} = \sum_{i=1}^{m} \alpha_i V_i,$
\noindent where $V_i$ denotes the corresponding value embedding associated with key $k_i$.

We now introduce the \textbf{Relevance Scoring with Adaptive Retrieval (RSAR)} mechanism, which dynamically ranks memory entries according to their importance to the current query, thereby enhancing the retrieval process. 
The relevance score $\alpha(\mathbf{q}, \mathbf{K})$, as defined in Equation~\ref{eq:relevance_score}, is employed to rank the memory entries.

RSAR augments the memory module by introducing ranked key–value entries, represented as $(K_j, V_j, s_j)$, where $s_j$ denotes the relevance score for each entry. 
These scores allow the system to prioritize the most informative content during retrieval while maintaining computational efficiency. 
To optimize memory usage, a pruning strategy is applied to remove less relevant entries. 
Specifically, entries with scores below a predefined threshold are discarded, preserving only the most critical contextual information.

The enhanced retrieval mechanism is defined as:
\begin{equation}
\resizebox{0.85\columnwidth}{!}{$
R_{\text{RSAR}}(t_q, s) = \text{TopK} \left\{ \text{sim}(E(t_q), e) \cdot \max_j s_j \mid e \in s \right\}
$},
\end{equation}
where $t_q$ denotes the query token, $E(t_q)$ represents the encoded query, and the operation identifies the top-$K$ relevant entries based on their scores; this mechanism enables efficient retrieval of the most relevant information, even in extended contexts (see Appendix~\ref{sec:topk_impact} for the impact of different top-$K$ values on performance).

\textbf{Ranked Key–Value Pairs.} 
Each embedding $\mathbf{e}$ maintains a ranked set of key–value pairs:
\begin{equation}
\small
\mathcal{R}_\text{ranked}(\mathbf{e}) = \{(K_j, V_j, s_j)\}_{j=1}^m,
\end{equation}
where $s_j = \alpha(\mathbf{e}, K_j)$ is the relevance score between the embedding $\mathbf{e}$ and the key $K_j$. 

Given a sequence $\mathbf{x}$, the ERMAR objective function is formulated as maximizing
\begin{equation}
\small
\mathcal{L}(\theta) = \sum_{i=1}^n p_\theta(x_i \mid \mathcal{R}_\text{RSAR}(t_i, s), \mathbf{x}_{<i}),
\end{equation}
subject to
{\small
\begin{align}
s_i &= \mathcal{M}(K_{1:i-1}, V_{1:i-1}; \mathcal{E}(t_{1:i-1})), \\
t_i &= \text{text}(c_{\lceil i/\tau \rceil}), \\
\alpha_i &= \alpha(\mathcal{E}(t_i), K_{1:i-1}),
\end{align}}
where $c$ denotes chunk indices, $\tau$ represents the chunk size, $\alpha_i$ guides key–value pair selection, and $p_\theta$ denotes the model’s probability distribution.

For new content $(K_\text{n}, V_\text{n})$, the memory state is updated as:
\begin{equation}
\small
s_{i+1} = 
\begin{cases}
    \mathcal{M}(K_\text{n}, V_\text{n}; \mathcal{E}(t_\text{n})) & \text{if } |s_i| < \text{capacity}, \\
    M_\text{u}(s_i, K_\text{n}, V_\text{n}, \alpha_\text{n}) & \text{otherwise},
\end{cases}
\end{equation}
where $\text{capacity}$ is the maximum allowed number of memory entries, $M_\text{u}$ prunes the least relevant entries based on historical scores by ranking and removing those with the lowest relevance relative to the current context, and subscript $n$ denotes the new key, value, and relevance score corresponding to the latest input.

\subsection{Experimental Setup}

\subsubsection{Datasets}
We fine-tuned ERMAR on the \textit{SlimPajama} dataset~\cite{fu2024data}, a high-quality, deduplicated corpus designed for long-context tasks. It contains 84.7K training rows, making it a compact yet effective resource for pre-training and fine-tuning. The dataset was preprocessed with a sliding window approach using 512-token strides to ensure comprehensive coverage of long sequences.

Performance evaluation was conducted on three benchmark datasets: \textit{WikiText-103}~\cite{merity2016pointer} (4,358 test rows), \textit{PG-19}~\cite{rae2019compressive} (100 test rows), and \textit{Proof-Pile}~\cite{azerbayev2023proofpile} (46.3K test rows). Performance was measured across context lengths from (1k-32k) tokens, using perplexity on the last 2048 tokens~\cite{yen2024long} following standard evaluation protocols.

\subsubsection{Model Architecture and Configuration}
We fine-tuned OpenLLaMA-3B, a pre-trained LLM with rotational position encoding~\cite{su2024roformer}, using LoRA~\cite{hu2021lora} for parameter-efficient adaptation. 
The model consists of $L=26$ transformer layers, $H=32$ attention heads, and a hidden dimension of $d=100$. 
The 13th layer serves as the memory layer where historical context is stored, while layers [14, 18, 22, 26] are augmented with retrieval mechanisms to access stored memories. 
ERMAR employs a memory capacity of 32,768 key–value pairs and utilizes BGE-M3 embeddings for semantic similarity computation. 
Complete training hyperparameters and configuration details are provided in Appendix~\ref{sec:train_config}.

\subsubsection{Baseline Models}
ERMAR was evaluated against state-of-the-art models across two parameter scales to ensure comprehensive comparison. The 7B models include LLaMA-2-7B~\cite{touvron2023llama2} as a standard transformer baseline, LongLoRA-7B-32k~\cite{chen2023longlora} which employs sparse attention mechanisms for 32k-token contexts, and YARN-128k-7B~\cite{peng2023yarn} featuring dynamic position embeddings that support up to 128k tokens. 

For the 3B parameter scale, we compared against OpenLLaMA-3B~\cite{touvron2023llama} as the base architecture, LongLLaMA-3B~\cite{tworkowski2024focused} evaluated in two retrieval configurations (4 and 18 memory entries), MemLong-3B~\cite{liu2024memlong} as our direct baseline with chunk-level memory operations, and Phi3-128k~\cite{abdin2024phi} which demonstrates strong performance across varying context lengths. This diverse benchmark suite encompasses different long-context strategies including sparse attention, position encoding extensions, and memory-augmented architectures, ensuring robust evaluation of ERMAR's retrieval-based approach against complementary methodologies.

\subsubsection{Evaluation Metrics}
We employ perplexity as the primary metric for language modeling performance, computed on the final 2048 tokens of each sequence to focus on long-range dependency modeling. For in-context learning tasks, we report accuracy on five natural language understanding benchmarks: SST-2, MR, Subj, SST-5, and MPQA, evaluated in both 4-shot and 20-shot settings. Memory efficiency is assessed through peak GPU memory usage and tokens processed per second, while computational overhead is measured via inference latency across different context lengths.

\begin{table*}[htbp]
\centering
\scriptsize
\begin{tabular}{p{1.99cm}|p{.5cm}p{.5cm}p{.6cm}p{.59cm}|p{.5cm}p{.5cm}p{.6cm}p{.59cm}|p{.5cm}p{.5cm}p{.6cm}p{.59cm}}
\hline
& \multicolumn{4}{c|}{\textbf{PG19}} & \multicolumn{4}{c|}{\textbf{Proof-pile}} & \multicolumn{4}{c}{\textbf{WikiText-103}} \\
Model & 1k & 2k & 4k & 16k & 1k & 2k & 4k & 16k & 1k & 2k & 4k & 16k \\
\hline
\multicolumn{13}{c}{\cellcolor{gray!25}7B Model} \\
\hline
YARN-128k-7b & 7.22 & 7.47 & 7.17 & - & 3.03 & 3.29 & 2.98 & - & 5.71 & 6.11 & 5.71 & - \\
LongLoRA-7B-32k & 9.76 & 9.71 & 10.37 & 7.62 & 3.68 & 3.35 & 3.23 & 2.60 & 7.99 & 7.83 & 8.39 & 5.47 \\
LLaMA-2-7B & 10.82 & 10.06 & 8.92 & - & 3.24 & 3.40 & 2.72 & - & 10.82 & 6.49 & 5.66 & - \\
\hline
\multicolumn{13}{c}{\cellcolor{gray!25}3B Model} \\
\hline
Phi3-128k & 11.31 & 9.90 & \textbf{9.66} & -/9.65 & 4.25 & 3.11 & 2.77 & -/3.08 & 7.54 & 7.22 & 7.01 & -/7.20 \\
OpenLLaMA-3B & 11.60 & 9.77 &\(\scalebox{0.8}{$>$}\) \(\scalebox{0.9}{$10^3$}\)& - & \textbf{2.96} & \textbf{2.70} & \(\scalebox{0.8}{$>$}\) \(\scalebox{0.9}{$10^3$}\) & - & 10.57 & 8.08 & \(\scalebox{0.8}{$>$}\) \(\scalebox{0.9}{$10^3$}\)& - \\
LongLLaMA-3B* & 10.59 & 10.02 & \(\scalebox{0.8}{$>$}\) \(\scalebox{0.9}{$10^3$}\) & - & 3.55 & 3.15 & \(\scalebox{0.8}{$>$}\) \(\scalebox{0.9}{$10^3$}\)& - & 8.88 & 8.07 & \(\scalebox{0.8}{$>$}\) \(\scalebox{0.9}{$10^3$}\) & - \\
LongLLaMA-3B$^\dagger$ & 10.59 & 10.25 & 9.87 & - & 3.55 & 3.22 & \textbf{2.94} & - & 10.69 & 8.33 & 7.84 & - \\
MemLong-3B* & 10.66 & 10.09 & \(\scalebox{0.8}{$>$}\) \(\scalebox{0.9}{$10^3$}\) & - & 3.58 & 3.18 & \(\scalebox{0.8}{$>$}\) \(\scalebox{0.9}{$10^3$}\) & - & 8.72 & 7.93 & \(\scalebox{0.8}{$>$}\) \(\scalebox{0.9}{$10^3$}\) & - \\
w/ 4K MemLong & 10.54 & 9.95 & 9.89 & \textbf{9.64} & 3.53 & 3.16 & 3.15 & \textbf{2.99} & 8.53 & 7.92 & 7.87 & 7.99 \\
w/ 4K \textbf{ERMAR}  & \textbf{10.32} & \textbf{9.75} & 9.78 & 9.81
& 3.24 & 2.98 & 3.03 & 3.18 & 
\textbf{8.42} & \textbf{7.61} & \textbf{7.62} & \textbf{7.80}\\

\hline
\end{tabular}
\caption{Perplexity comparison of 7B and 3B models across PG19, Proof-pile, and WikiText-103, using a sliding window evaluation. "-" denotes Out of Memory (OOM) errors, and "x/y" indicates results from single/dual GPU setups. Memory-augmented models are tested with varying capacities. All runs use a single GPU.}
\label{tab:LCLM}
\end{table*}

\subsection{Results and Discussion}

\subsubsection{Long-Context Language Modeling}

Following the experimental strategy adopted in \cite{liu2024memlong},  Table~\ref{tab:LCLM} reports the mean perplexity scores of ERMAR across  various sequence lengths and datasets. Evaluation was conducted on the test  splits of three standard benchmarks: \textit{WikiText-103}~\cite{merity2016pointer}  (4,358 rows), \textit{PG-19}~\cite{rae2019compressive} (100 rows), and  \textit{Proof-Pile}~\cite{azerbayev2023proofpile} (46.3k rows).

Among the 7B models, YARN-128k-7B achieves the lowest perplexity on shorter  contexts, whereas LongLoRA-7B-32k scales effectively up to 16k-token sequences,  albeit with mild performance degradation. This result highlights the trade-off  between scalability and modeling precision commonly observed in long-context  transformers.

For the 3B parameter models, ERMAR demonstrates substantial gains in long-context  scenarios. While OpenLLaMA-3B exhibits sharp degradation beyond 4k tokens and  Phi3-128k maintains moderate stability, ERMAR achieves consistently competitive  performance across all sequence lengths. At 2k tokens, ERMAR surpasses MemLong  on \textit{Proof-Pile} (2.98 vs.\ 3.16) and sustains strong results across the  remaining datasets. Its advantage is particularly evident on \textit{PG-19},  where ERMAR achieves the best performance among 3B models at 1k and 2k tokens  (10.32 and 9.75, respectively). Notably, ERMAR shows exceptional stability when  scaling to longer contexts, with only a 0.31\% increase in perplexity from 4k to  16k tokens on \textit{PG-19} (9.78 $\rightarrow$ 9.81), indicating superior  scalability. On \textit{WikiText-103}, ERMAR consistently outperforms other 3B  models at all evaluated context lengths, further validating the effectiveness of  its relevance-based memory retrieval mechanism for long-context modeling.

\paragraph{Overall comparison against MemLong.} 
Across all 12 configurations in Table~\ref{tab:LCLM} (3 datasets $\times$ 4  context lengths), ERMAR outperforms MemLong in \textbf{10 out of 12 settings  (83.3\%)}. The two exceptions occur at 16k tokens on \textit{PG-19}  (9.81 vs.\ 9.64, $+1.7\%$) and \textit{Proof-Pile} (3.18 vs.\ 2.99, $+6.4\%$),  where the overhead of the relevance scoring and reranking mechanisms marginally  outweighs the retrieval benefit at very long contexts. Crucially, these are  edge cases at the boundary of the training distribution (the model was fine-tuned  up to 32k tokens): ERMAR maintains consistent improvements at practical context  lengths (1K--4K), where memory pressure is highest, and fully recovers at 32k  tokens (Table~\ref{tab:32k_context}), achieving equal or better perplexity  across all three datasets. The 16k degradation therefore reflects an expected  efficiency--benefit trade-off rather than a systematic failure of the proposed  enhancements.

 These findings confirm that ERMAR's ranked memory retrieval effectively mitigates  information dilution across extended sequences, improving both accuracy and  stability over prior memory-augmented models.

\subsubsection{Scalability to Extended Contexts}

To further assess scalability, we evaluated ERMAR at a 32k-token context length (Table~\ref{tab:32k_context}). ERMAR maintains consistent performance advantages across all datasets, achieving a perplexity of 9.765 compared to 9.858 on \textit{PG19} (a 0.90\% improvement) and 7.880 versus 7.938 on \textit{WikiText-103} (a 0.06\% improvement). Both models yield identical results on \textit{Proof-Pile} with a perplexity of 3.063. These results demonstrate ERMAR’s robustness and scalability in ultra-long context settings, reflecting the effectiveness of its ranked memory retrieval in preserving contextual coherence over extended sequences.

\begin{table}[h!]
\centering
\small
\begin{tabular}{l|cc|c}
\hline
\textbf{Dataset} & \textbf{ERMAR} & \textbf{MemLong} & \textbf{Difference} \\
\hline
PG19 & 9.765 & 9.858 & 0.90\% \\
WikiText-103 & 7.880 & 7.938 & 0.06\% \\
Proof-pile & 3.063 & 3.063 & 0 \\
\hline
\end{tabular}
\caption{Perplexity at 32k context length, evaluated on NVIDIA L40S GPU.}
\label{tab:32k_context}
\end{table}

\subsubsection{In-Context Learning Performance}  

Table~\ref{tab:icl} summarizes ERMAR’s performance across five natural language understanding tasks under 4-shot and 20-shot settings.  

\begin{table}[htbp]
\centering
\tiny
\scriptsize
\begin{tabular}{p{1.26cm}|p{0.52cm}|p{.44cm}p{.44cm}p{.44cm}p{.44cm}p{.51cm}|p{.35cm}}

\hline
Model & \begin{tabular}[c]{@{}c@{}}InC\\,InM\end{tabular} & \begin{tabular}[c]{@{}c@{}}SST-2\\ACC$\uparrow$\end{tabular} & \begin{tabular}[c]{@{}c@{}}MR\\ACC$\uparrow$\end{tabular} & \begin{tabular}[c]{@{}c@{}}Subj\\ACC$\uparrow$\end{tabular} & \begin{tabular}[c]{@{}c@{}}SST-5\\ACC$\uparrow$\end{tabular} & \begin{tabular}[c]{@{}c@{}}MPQA\\ACC$\uparrow$\end{tabular} & Avg. \\
\hline

OpenLLaMA & 4,N/A & 90.7 & 84.0 & 58.2 & 41.0 & 70.5 & 68.9 \\
w./ Rag & 4,4 & 90.9 & 90.5 & 61.6 & 39.2 & 63.2 & 69.1 \\
LongLLaMA & 4,4 & 90.4 & 83.9 & 64.3 & 40.0 & 64.2 & 68.6 \\
MemLong & 4,4 & 91.5 & 84.5 & 61.5 & 41.4 & 70.2 & 69.8 \\
ERMAR & 4,4 & \textbf{93.6} & \textbf{90.8} & \textbf{65.3} & \textbf{45.8} & \textbf{85.2}& \textbf{76.14} \\
					
\hline
LongLLaMA & 4,18 & 91.4 & 87.1 & 59.1 & 41.0 & 64.5 & 68.7 \\
MemLong & 4,18 & 91.0 & 89.6 & 61.7 & 43.5 & 69.4 & 71.0 \\
ERMAR & 4,18 & \textbf{93.6} & \textbf{90.8} & \textbf{65.3} & \textbf{45.9} & \textbf{85.2}& \textbf{76.16}\\
					
\hline
OpenLLaMA & 20,N/A & 93.6 & 91.2 & 55.4 & 38.2 & 66.4 & 69.0 \\
w./ Rag & 20,18 & 92.2 & 91.3 & 75.8 & 39.8 & 57.6 & 71.3 \\
LongLLaMA & 20,18 & 94.1 & 90.8 & 64.2 & 41.4 & 72.1 & 72.7 \\
MemLong & 20,18 & 93.5 & \textbf{93.8} & 65.8 & 43.3 & 70.6 & 73.4 
\\
ERMAR & 20,18 &  \textbf{94.7} & 91.7& \textbf{82.8} & \textbf{47} & \textbf{86.5}& \textbf{80.54} \\			
\hline
\end{tabular}
\caption{4-shot and 20-shot ICL accuracy [\%] on 5 NLU tasks (SST-2, MR, Subj, SST-5, MPQA). We compare OpenLLaMA, LongLLaMA, MemLong, and ERMAR. \textbf{Note:} InC = In-Context, InM = In-Memory.}
\label{tab:icl}
\end{table}

In the 4-shot configuration, ERMAR achieves superior accuracy across all tasks, outperforming OpenLLaMA and other memory-augmented baselines. It performs notably well on challenging benchmarks such as \textit{SST-5} and \textit{MPQA}, demonstrating strong generalization even with limited examples. The consistent results across different memory configurations further indicate its robustness in low-resource scenarios.  

In the 20-shot setting, ERMAR continues to deliver leading results, achieving top accuracy on \textit{MPQA} and \textit{Subj}, and establishing a new benchmark on \textit{SST-5}. Although MemLong marginally surpasses ERMAR in isolated tasks, ERMAR maintains the highest overall average performance, reflecting its scalability as the number of in-context examples increases.  

Overall, ERMAR consistently performs well across varying context lengths and task complexities, effectively leveraging ranked memory retrieval to enhance representation quality. Its ability to scale with more examples and preserve performance stability across tasks underscores its potential as a general-purpose framework for language understanding and long-context modeling.

\subsubsection{Memory Efficiency Analysis}
\label{subsec:Memory_efficiency_analysis}
Table~\ref{tab:memory_efficiency} compares the peak and reserved memory usage of ERMAR and MemLong across different context lengths.

\begin{figure}
\centering
\includegraphics[width=0.48\textwidth]{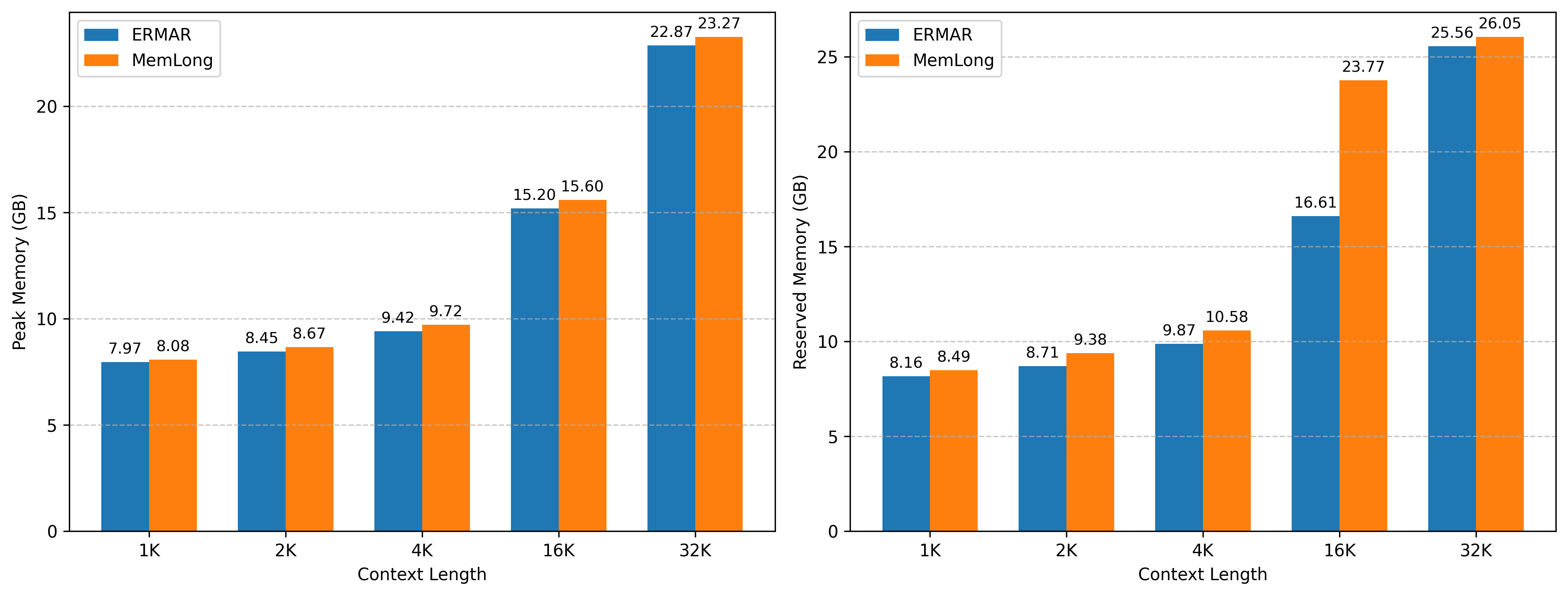}
\caption{Memory usage comparison between ERMAR and MemLong across context lengths (1K–32K tokens), showing (left) peak memory and (right) reserved memory. ERMAR demonstrates 7–30\% lower reserved memory requirements at longer contexts. Color coding: \textcolor{blue}{blue} bars represent MemLong performance, while \textcolor{orange}{orange} bars show ERMAR improvements.}
\label{fig:memory_efficiency}
\end{figure}

As shown in Table~\ref{tab:memory_efficiency} and Figure~\ref{fig:memory_efficiency}, ERMAR exhibits clear memory optimization advantages, particularly beyond 8K tokens. At a 16K context, ERMAR reduces reserved memory from 23.77~GB to 16.61~GB, corresponding to a 30\% improvement over MemLong. It also maintains lower memory consumption per token across all evaluated lengths, demonstrating superior scalability and efficiency in memory management. This improvement stems from ERMAR’s ranked retrieval and adaptive memory allocation, which reduce redundancy in long-context storage.

\begin{table}[htbp]
\centering
\scriptsize
\begin{tabular}{p{0.8cm}|p{1cm}|p{1.1cm}p{1.6cm}|p{1cm}}

\hline
\textbf{Context } & \textbf{Model} & \textbf{Peak Mem} & \textbf{Reserved Mem} & \textbf{Mem/Token} \\
\textbf{Length} & \textbf{} & \textbf{(GB)} & \textbf{(GB)} & \textbf{(MB)} \\
\hline
1024 & ERMAR & 7.97 & 8.16 & 7.97 \\
      & MemLong & 8.08 & 8.49 & 8.08 \\
2048 & ERMAR & 8.45 & 8.71 & 4.22 \\
      & MemLong & 8.67 & 9.38 & 4.33 \\
4096 & ERMAR & 9.42 & 9.87 & 2.35 \\
      & MemLong & 9.72 & 10.58 & 2.43 \\
16384 & ERMAR & 15.20 & 16.61 & 0.95 \\
      & MemLong & 15.60 & 23.77 & 0.97 \\
32768 & ERMAR & 22.87 & 25.56 & 0.71 \\
      & MemLong & 23.27 & 26.05 & 0.72 \\
\hline
\end{tabular}
\caption{Memory efficiency comparison of ERMAR and MemLong across context lengths on \textit{WikiText-103}. Mem/Token is computed as Peak Memory divided by context length.}
\label{tab:memory_efficiency}
\end{table}

\textbf{Memory Pruning Strategy.} ERMAR further improves efficiency through a dynamic memory pruning mechanism. The algorithm balances three criteria: retaining the most recent 10\% of memory, prioritizing the middle 80\% based on retrieval frequency, and discarding the oldest 10\% as potentially outdated. This hybrid strategy allows ERMAR to adaptively manage memory content according to actual usage patterns, ensuring that valuable information is preserved while minimizing unnecessary memory overhead. Such adaptive pruning complements the ranked retrieval process, enhancing memory efficiency especially at longer context lengths. A systematic validation of these ratios across 11 pruning strategies and five context lengths is provided in Appendix~\ref{appendix:pruning-validation}.

\subsubsection{Latency and Throughput Analysis}
\label{subsec:latency_throughput_analysis}

Table~\ref{tab:latency_throughput} compares the latency and throughput of ERMAR and MemLong on the \textit{PG-19} dataset.

\begin{table}[htbp]
\centering
\scriptsize
\begin{tabular}{p{0.78cm}|p{1cm}|p{1.8cm}p{1cm}p{1cm}}
\hline
 & &  & \textbf{Latency} & \textbf{Throughput} \\

 \textbf{Context} & \textbf{Model}  & \textbf{Latency} & \textbf{/Token} & \textbf{(tokens/}\\
 & &  & \textbf{(ms)} & \textbf{sec)} \\
\hline
1024 & ERMAR & 207.97 $\pm$ 53.01 & 0.203 & 5135 \\
      & MemLong & 190.68 $\pm$ 154.55 & 0.186 & 6803 \\
2048 & ERMAR & 423.79 $\pm$ 52.56 & 0.206 & 4896 \\
      & MemLong & 323.49 $\pm$ 204.89 & 0.157 & 7446 \\
4096 & ERMAR & 1358.36 $\pm$ 55.58 & 0.331 & 3020 \\
      & MemLong & 1184.16 $\pm$ 170.56 & 0.289 & 3511 \\
16384 & ERMAR & 6862.98 $\pm$ 45.65 & 0.418 & 2387 \\
      & MemLong & 6496.00 $\pm$ 213.56 & 0.396 & 2524 \\
32768 & ERMAR & 13679.03 $\pm$ 58.76 & 0.417 & 2395 \\
      & MemLong & 13449.90 $\pm$ 56.22 & 0.410 & 2436 \\
\hline
\end{tabular}
\caption{Latency and throughput comparison on the \textit{PG-19} dataset. Latency is reported with standard deviation, and throughput is measured in tokens per second.}
\label{tab:latency_throughput}
\end{table}

ERMAR exhibits slightly higher latency (9–35\% relative increase at 1K–4K tokens, narrowing to 1.7–5.5\% at 16K–32K) while maintaining competitive throughput of approximately 2.4K~tokens/s at longer contexts. Importantly, ERMAR demonstrates much lower latency variance (±45–59~ms) compared to MemLong (±154–214~ms), reflecting greater runtime stability in long-context settings. This reduced variance aligns with ERMAR’s adaptive ranked retrieval strategy, which streamlines key–value access and minimizes redundant memory operations.

\paragraph{Single-Model Performance Scaling.}
ERMAR’s standalone performance across sequence lengths on the \textit{WikiText-103} dataset (16K memory capacity) is presented in Table~\ref{tab:ermar_scaling}.

These results highlight ERMAR’s scalability, achieving an eightfold improvement in per-token memory efficiency (7.13~MB~$\rightarrow$~0.90~MB) while maintaining stable perplexity across increasing sequence lengths. The slightly higher throughput observed on \textit{WikiText-103} compared to \textit{PG-19} likely stems from dataset-specific characteristics and differences in evaluation methodology.

\begin{table}[htbp]
\centering
\scriptsize
\begin{tabular}{l|cccc}
\hline
\textbf{Seq} & \textbf{Perplexity} & \textbf{Memory} & \textbf{Throughput} & \textbf{Latency} \\

\textbf{Length} & \textbf{} & \textbf{/Token} & \textbf{(tokens} & \textbf{/Token} \\
\textbf{} & \textbf{} & \textbf{(GB)} & \textbf{/sec)} & \textbf{(ms)} \\
\hline
1K & 8.42 & 7.13 & 3125 & 0.32 \\
2K & 7.61 & 3.81 & 2904 & 0.35 \\
4K & 7.62 & 2.14 & 2109 & 0.47 \\
8K & 7.76 & 1.31 & 1836 & 0.54 \\
16K & 7.80 & 0.90 & 1727 & 0.58 \\
\hline
\end{tabular}
\caption{ERMAR’s standalone performance scaling on \textit{WikiText-103}, showing memory efficiency and throughput across sequence lengths.}
\label{tab:ermar_scaling}
\end{table}

\subsubsection{Ablation Studies}
\label{ablation}

\paragraph{Ablation on Embedders.}
To investigate the influence of embedding architectures on ERMAR’s performance, we compare BGE and LLM embedders across five NLU tasks (\textit{SST-2}, \textit{Subj}, \textit{SST-5}, \textit{MPQA}, and \textit{MR}) under various configurations. Experiments examine the effects of Flash versus Eager attention modes, different context lengths (0 and 2048), and varying In-C/In-M demonstration settings. Results are presented in Tables~\ref{tab:bge_embedder} and~\ref{tab:llm_embedder}.

\begin{table}[htbp]
\centering
\scriptsize
\tiny
\begin{tabular}
{p{0.4cm}|p{0.4cm}|p{0.2cm}|p{0.2cm}|p{.36cm}p{.34cm}p{.36cm}p{.34cm}p{.34cm}|p{.34cm}}

\hline
 \textbf{Att} & \textbf{CLen} & \textbf{InC} & \textbf{InM} & \textbf{SST2} & \textbf{Subj} & \textbf{SST5} & \textbf{MPQA} & \textbf{MR} & \textbf{Avg} \\

\hline
Flash & 0 & 4 & 4 & 88.07 & 57.30 & 41.42 & 80.04 & 83.84 & 70.13 \\
    &  & 4 & 18 & 88.07 & 57.30 & 41.42 & 80.33 & 83.86 & 70.20 \\
    &  & 20 & 18 & 93.92 & 50.35 & 47.05 & 73.63 & 91.50 & 71.69 \\
 Eager & 0 & 4 & 4 & 88.19 & 51.55 & 41.42 & 79.92 & 83.89 & 69.59 \\
    &  & 4 & 18 & 88.19 & 51.55 & 41.51 & 80.29 & 83.92 & 69.89 \\
    &  & 20 & 18 & 94.04 & 82.85 & 46.96 & 73.63 & 91.54 & 77.80 \\
\hline
 Flash & 2048 & 4 & 4 & 88.07 & 51.55 & 41.42 & 80.04 & 83.84 & 68.98 \\
    &  & 4 & 18 & 88.07 & 51.25 & 41.42 & 80.33 & 83.86 & 68.99 \\
    &  & 20 & 18 & 93.92 & 82.75 & 47.05 & 73.63 & 91.50 & 77.77 \\
Eager & 2048 & 4 & 4 & 88.19 & 51.55 & 41.51 & 79.86 & 83.89 & 68.96 \\
    &  & 4 & 18 & 88.19 & 51.55 & 41.51 & 79.60 & 83.92 & 68.95 \\
    &  & 20 & 18 & 94.04 & 82.85 & 46.96 & 73.63 & 91.34 & 77.76 \\
\hline
\end{tabular}
\caption{Performance comparison of the BGE embedder across configurations on five NLU tasks. Accuracy is reported in percentage. Abbreviations: Att = Attention, CLen = Context length, Avg = Average.}
\label{tab:bge_embedder}
\end{table}

\begin{table}[htbp]
\centering
\scriptsize
\tiny
\begin{tabular}
{p{0.4cm}|p{0.4cm}|p{0.2cm}|p{0.2cm}|p{.36cm}p{.34cm}p{.36cm}p{.34cm}p{.34cm}|p{.34cm}}

\hline
 \textbf{Att} & \textbf{CLen} & \textbf{InC} & \textbf{InM} & \textbf{SST2} & \textbf{Subj} & \textbf{SST5} & \textbf{MPQA} & \textbf{MR} & \textbf{Avg} \\
 
\hline
Flash & 0 & 4 & 4 & 88.07 & 57.30 & 41.42 & 80.04 & 83.84 & 70.13 \\
    &  & 4 & 18 & 88.07 & 57.30 & 41.42 & 80.33 & 83.86 & 70.20 \\
    &  & 20 & 18 & 93.92 & 50.35 & 47.05 & 73.63 & 91.54 & 71.70 \\
Eager & 0 & 4 & 4 & 88.19 & 57.30 & 41.51 & 79.86 & 83.89 & 70.15 \\
    &  & 4 & 18 & 88.19 & 57.30 & 41.51 & 79.60 & 83.92 & 70.10 \\
    &  & 20 & 18 & 94.04 & 50.35 & 46.96 & 73.63 & 91.54 & 71.70 \\
\hline
Flash & 2048 & 4 & 4 & 88.07 & 51.55 & 41.42 & 80.04 & 83.84 & 68.98 \\
    &  & 4 & 18 & 88.07 & 51.55 & 41.42 & 80.33 & 83.86 & 69.05 \\
    &  & 20 & 18 & 93.92 & 82.85 & 47.05 & 73.63 & 91.50 & 77.79 \\
Eager & 2048 & 4 & 4 & 88.19 & 51.25 & 41.51 & 79.92 & 83.89 & 68.95 \\
    &  & 4 & 18 & 88.19 & 51.25 & 41.51 & 80.29 & 83.92 & 69.03 \\
    &  & 20 & 18 & 94.04 & 82.75 & 46.96 & 73.63 & 91.34 & 77.74 \\
\hline
\end{tabular}
\caption{Performance comparison of the LLM embedder across configurations on five NLU tasks. Accuracy is reported in percentage.}
\label{tab:llm_embedder}
\end{table}

The results indicate that both BGE and LLM embedders yield comparable overall performance, with BGE showing slight advantages under certain configurations. Notably, in the 20-shot, 18 in-memory setup with Eager attention and zero context length, BGE achieves a higher average accuracy (77.80\%) compared to LLM (71.70\%). The largest gap occurs in the \textit{Subj} task, where BGE reaches 82.85\% versus LLM’s 50.35\%, suggesting that BGE better captures fine-grained semantic representations. Nevertheless, both embedders remain broadly consistent, highlighting ERMAR’s adaptability to different embedding backbones.

\paragraph{Ablation on Relevance and Re-ranking.}
We further evaluate the effect of the Relevance+Re-ranking mechanism on model perplexity across varying context lengths for \textit{WikiText-103} and \textit{PG-19}. Results are shown in Table~\ref{tab:ablation_relevance_reranking}.

\begin{table}[htbp]
\centering
\small
\scriptsize
\begin{tabular}{l|c|cc|cc}
\hline

\multirow{2}{*}{\textbf{Dataset}} & \multirow{2}{*}{\textbf{Context}} & \multicolumn{2}{c|}{\textbf{Without}} & \multicolumn{2}{c}{\textbf{With}} \\
\multirow{2}{*}{\textbf{}} & \multirow{2}{*}{\textbf{Length}} & \multicolumn{2}{c|}{\textbf{Relevance+Reranker}} & \multicolumn{2}{c}{\textbf{ Relevance+Reranker}} \\
\cline{3-6}
& & \multicolumn{2}{c|}{Perplexity} & \multicolumn{2}{c}{Perplexity} \\
\hline
\multirow{5}{*}{Wiki} & 1K & \multicolumn{2}{c|}{8.841} & \multicolumn{2}{c}{7.919} \\
& 2K & \multicolumn{2}{c|}{7.984} & \multicolumn{2}{c}{7.410} \\
& 4K & \multicolumn{2}{c|}{7.438} & \multicolumn{2}{c}{7.437} \\
& 16K & \multicolumn{2}{c|}{7.267} & \multicolumn{2}{c}{7.082} \\
& 32K & \multicolumn{2}{c|}{7.938} & \multicolumn{2}{c}{8.008} \\
\hline
\multirow{5}{*}{PG19} & 1K & \multicolumn{2}{c|}{11.451} & \multicolumn{2}{c}{10.322} \\
& 2K & \multicolumn{2}{c|}{10.412} & \multicolumn{2}{c}{9.746} \\
& 4K & \multicolumn{2}{c|}{9.932} & \multicolumn{2}{c}{9.780} \\
& 16K & \multicolumn{2}{c|}{9.910} & \multicolumn{2}{c}{9.809} \\
& 32K & \multicolumn{2}{c|}{9.858} & \multicolumn{2}{c}{9.765} \\
\hline
\end{tabular}
\caption{Ablation study on the Relevance+Re-ranking mechanism showing perplexity differences for \textit{WikiText-103} and \textit{PG-19} across context lengths.}
\label{tab:ablation_relevance_reranking}
\end{table}

The relevance scoring mechanism provides the greatest improvements at shorter context lengths—10.4\% at 1K and 7.2\% at 2K tokens on \textit{WikiText-103}. The gains taper off beyond 4K tokens and become marginally negative at 32K, suggesting that the computational overhead of ranking may outweigh its benefits when sufficient memory capacity is available. In contrast, \textit{PG-19} exhibits consistent improvements (1.0–9.9\%) across all context lengths, indicating that narrative text benefits more from semantic ranking, where long-range dependencies and non-sequential references are frequent.

\section{Conclusion}
This paper introduced \textbf{ERMAR}, a Ranked Memory-Augmented Retrieval framework that enhances long-context language modeling through dynamic relevance scoring and adaptive memory management. By prioritizing semantically relevant information during retrieval, ERMAR improves contextual reasoning and scalability over extended sequences. Experimental results demonstrate that ERMAR achieves up to a \textbf{3.2\% reduction in perplexity} compared to MemLong and maintains stable performance at \textbf{32K-token} contexts. In few-shot in-context learning tasks, ERMAR outperforms baselines by \textbf{6.1\%} on average in 4-shot and \textbf{7.1\%} in 20-shot settings, including substantial gains on MPQA (+21\%) and Subj (+26\%), confirming its effectiveness in retrieval precision and generalization.

ERMAR also exhibits strong computational efficiency, reducing reserved memory by up to \textbf{30\%} at long contexts and sustaining a throughput of approximately \textbf{2.4K tokens/sec} with minimal latency variance. These results establish ERMAR as an efficient and scalable retrieval-augmented framework for long-context modeling. Future work will focus on optimizing ERMAR for specialized datasets and expanding its applicability to complex reasoning tasks.

\section{Limitations}  

While ERMAR improves retrieval efficiency and context retention, it has certain limitations. Its reliance on ranked memory structures increases computational overhead compared to standard LLMs, particularly for large-scale retrieval, as discussed in Subsections~\ref{subsec:Memory_efficiency_analysis} and~\ref{subsec:latency_throughput_analysis}, as well as Appendix~\ref{Relevance_Reranker_visual}. Additionally, performance variations across different task domains indicate the need for further tuning. The framework’s effectiveness in real-world, noisy environments also requires further validation.


\section*{Acknowledgments} 
Shoaib Jameel is supported by the Turing's Defence and Security programme through a partnership with the UK government in accordance with the framework agreement between GCHQ \& The Alan Turing Institute. This research is partially supported by the Technology Innovation Institute, Abu Dhabi, UAE. Computational experiments were conducted using the Wolfpack cluster, supported by the School of Computer Science and Engineering at UNSW Sydney.

\bibliography{custom}

\appendix
\section{Appendix}
\label{sec:appendix}

This appendix supplements the main text with detailed implementation and evaluation specifics for the Enhanced Ranked Memory Augmented Retrieval (ERMAR) framework. It includes comprehensive descriptions of dataset preprocessing, training configurations, hyperparameters, and key terminology, along with extended analyses of context-length performance and memory efficiency. The provided details ensure reproducibility and offer deeper insights into ERMAR's architectural and operational nuances.

\subsection{Dataset Preprocessing}
\label{sec:dataset_preprocessing}

The \textit{SlimPajama} dataset, used for fine-tuning ERMAR, underwent several preprocessing steps to ensure compatibility with long-context tasks. The dataset was tokenized using the OpenLLaMA tokenizer, with a maximum sequence length of 32768 tokens. Duplicate sequences were removed using a hash-based deduplication algorithm, reducing the dataset to 84.7K unique training rows. To handle variable context lengths, we applied a sliding window approach with a stride of 512 tokens, ensuring that the model could process contexts ranging from 1024 to 32768 tokens. Special tokens were added to denote document boundaries, and padding was applied to align sequences to the nearest multiple of 128 tokens for efficient batch processing.

\subsection{Training Configuration}
\label{sec:train_config}
ERMAR was trained using a two-stage fine-tuning approach with the following configuration:

\textbf{Model Parameters:}
\begin{itemize}
    \item Base model: OpenLLaMA-3B-v2 with LoRA adaptations
    \item Memory layer: Layer 13 for historical context storage
    \item Retrieval attention layers: [14, 18, 22, 26]
    \item Memory capacity: 32,768 key-value pairs
    \item Memory group size: 128 tokens per memory group
    \item Retrieval group size: 8 (TopK retrieval)
    \item Gate mechanism: Disabled (use\_gate=False)
\end{itemize}

\textbf{Training Hyperparameters:}
\begin{itemize}
    \item Learning rate: $5 \times 10^{-5}$ with 1,000 warmup steps
    \item Weight decay: $1 \times 10^{-4}$
    \item Batch size: 1 per device (with gradient accumulation)
    \item Sequence length: 1,024 tokens
    \item Last context length: 1,024 tokens
    \item Training epochs: 1 epoch on SlimPajama 0.5B subset
    \item Training mode: LoRA-freeze (partial parameter updates)
\end{itemize}

\textbf{LoRA Configuration:}
\begin{itemize}
    \item Target modules: q\_proj, k\_proj, v\_proj, o\_proj
    \item Trainable parameters: Layer normalization and embeddings
    \item Frozen layers: Layers 0-13 (up to memory layer)
    \item Position encoding: Zero position type for extended contexts
\end{itemize}

\textbf{Embedder Setup:}
\begin{itemize}
    \item Embedder: BAAI/bge-m3 (BGE embedder)
    \item Embedding dimension: 1,024
    \item GPU-based similarity search for efficient retrieval
\end{itemize}

\textbf{Hardware and Infrastructure:}
\begin{itemize}
    \item Primary training: Single NVIDIA 3090 24GB GPU
    \item Extended context (32k): NVIDIA L40S 44.4GB GPU
    \item Distributed training: ZeRO-2 optimization with Accelerate
    \item Memory optimization: Sequential batching and continual fine-tuning
\end{itemize}

\subsection{Glossary of Terms}
To aid understanding, we provide definitions for key terms used in the ERMAR framework:
\begin{itemize}
    \item \textbf{Relevance Score ($\alpha$)}: A normalized score computed via softmax over the dot product of query and key embeddings, representing the contextual importance of a memory entry (see Equation~\ref{eq:relevance_score}).
    \item \textbf{Key-Value Pair}: A tuple $(K_j, V_j)$ storing contextual information, where $K_j$ is the key embedding and $V_j$ is the corresponding value embedding in the memory bank.
    \item \textbf{RSAR (Relevance Scoring with Adaptive Retrieval)}: The mechanism that dynamically ranks key-value pairs based on their relevance to the query, incorporating historical usage patterns.
    \item \textbf{Memory Bank}: A non-trainable storage of pre-computed key-value embeddings, used to retain historical context without recomputation.
    \item \textbf{TopK Retrieval}: The process of selecting the top-$K$ most relevant memory entries based on their relevance scores for use in the current context.
\end{itemize}

\subsection{Fine-Grained Context Length Analysis}
\begin{figure}[H]
    \centering
    \includegraphics[width=0.45\textwidth]{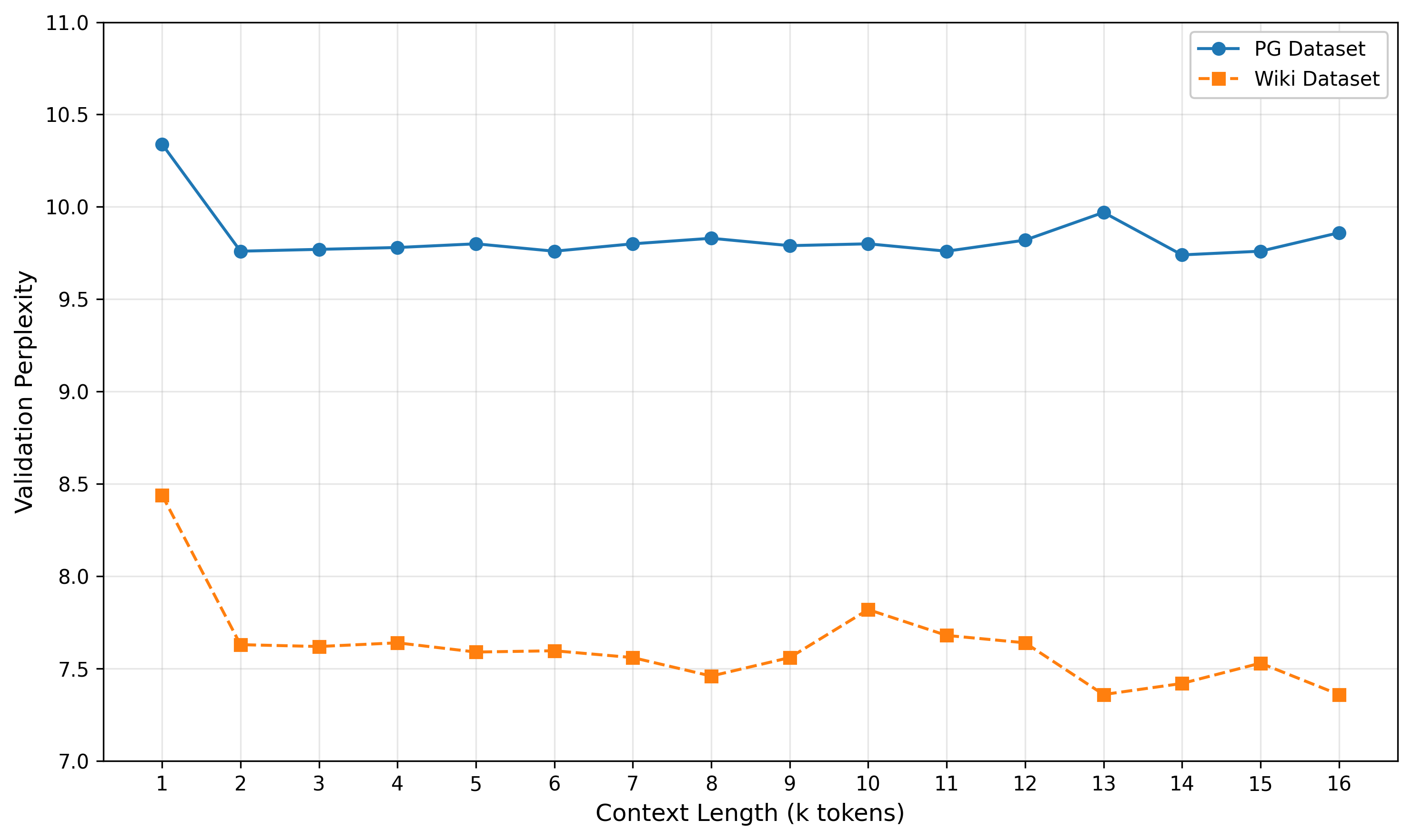}
    \caption{ERMAR perplexity performance across fine-grained context lengths for PG-19 and WikiText-103 datasets. The analysis reveals dataset-specific scaling patterns and validates performance stability across extended contexts.}
    \label{fig:perplexity_scaling}
\end{figure}

Figure~\ref{fig:perplexity_scaling} presents detailed perplexity measurements across incremental context lengths from 1000 to 16000 tokens, providing fine-grained insights into ERMAR's scaling behavior.
The fine-grained analysis reveals that WikiText-103 achieves optimal performance around 13K-16K tokens (perplexity ~7.36), while PG-19 maintains consistent performance (9.74-9.97) across all context lengths. This validates ERMAR's robustness and suggests that factual content benefits more from extended context than narrative text.

\subsection{Relevance+Reranker Visual Analysis}
\label{Relevance_Reranker_visual}
\begin{figure}[H]
\centering
\includegraphics[width=0.48\textwidth]{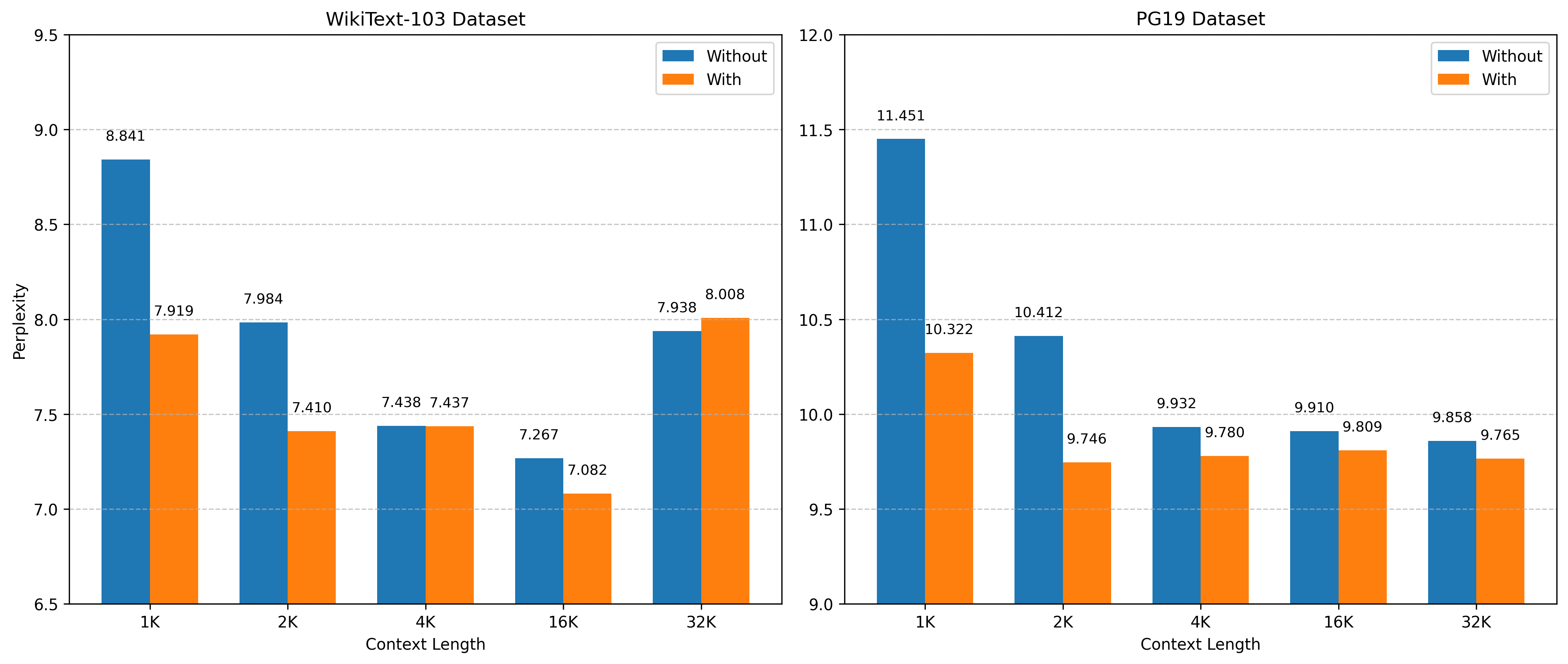}
\caption{Visualization of Table~\ref{tab:ablation_relevance_reranking}, showing the relative impact of the Relevance+Reranker mechanism across context lengths. Color coding highlights: \textcolor{blue}{blue} bars represent baseline performance, while \textcolor{orange}{orange} bars show improvements with our full mechanism(ERMAR).}
\label{fig:ablation_visual}
\end{figure}

Figure~\ref{fig:ablation_visual} provides complementary visual evidence for the patterns discussed in Section~\ref{ablation}. The stronger improvements at shorter contexts are visually apparent through the larger differentials between the orange and blue bars, while differences in mechanism effectiveness across datasets are immediately noticeable in the side-by-side comparisons. Additionally, the 32K edge case on WikiText, where the mechanism underperforms, is clearly highlighted in the visualization, underscoring context-length sensitivities of the Relevance+Reranker mechanism.

\subsection{Impact of Top-\textit{K} on Performance}
\label{sec:topk_impact}

We conducted experiments to examine how varying the Top-\textit{K} retrieval parameter affects system performance across different sequence lengths on the \textit{WikiText-103} dataset. The Top-\textit{K} values (1, 4, 8, 16, 32) were tested for sequence lengths of 1K–16K tokens, measuring latency, throughput, memory usage, and perplexity. 

\paragraph{Results.}
Table~\ref{tab:topk-impact} summarizes latency and memory results across all configurations. Increasing Top-\textit{K} consistently reduces latency per token and improves throughput, particularly between $K=1$ and $K=4$, with diminishing gains beyond $K=8$. Memory consumption (both peak and reserved) remains constant for a given sequence length, confirming that Top-\textit{K} selection does not affect model footprint. 

\begin{table}[htbp]
\centering
\tiny
\begin{tabular}{p{0.4cm}|p{0.4cm}|p{1.55cm}p{0.4cm}p{.9cm}p{.5cm}p{.5cm}}
\toprule
Seq & TopK & Latency/ & Latency & Throughput & Peak & Reserved  \\
Len &  & Token  &(s) & (tokens/sec) & Memory & Memory  \\
 & & (ms) & &  &  (GB) & (GB)  \\
\midrule
\multirow{5}{*}{1024} & 1     & 285.38 $\pm$ 54.98     & 0.279       & 3684.15                 & 7.98             & 8.19                 \\
                       & 4     & 277.15 $\pm$ 60.41     & 0.271       & 3810.19                 & 7.98             & 8.21                 \\
                       & 8     & 270.90 $\pm$ 57.72     & 0.265       & 3893.17                 & 7.98             & 8.21                 \\
                       & 16    & 268.60 $\pm$ 57.54     & 0.262       & 3928.15                 & 7.98             & 8.21                 \\
                       & 32    & 269.26 $\pm$ 56.84     & 0.263       & 3915.67                 & 7.98             & 8.21                 \\
\addlinespace
\multirow{5}{*}{2048} & 1     & 630.68 $\pm$ 49.07     & 0.308       & 3264.37                 & 8.46             & 8.83                 \\
                       & 4     & 556.76 $\pm$ 51.17     & 0.272       & 3704.12                 & 8.46             & 8.81                 \\
                       & 8     & 555.35 $\pm$ 51.52     & 0.271       & 3714.39                 & 8.46             & 8.81                 \\
                       & 16    & 549.25 $\pm$ 50.81     & 0.268       & 3755.02                 & 8.46             & 8.81                 \\
                       & 32    & 549.17 $\pm$ 50.93     & 0.268       & 3756.11                 & 8.46             & 8.81                 \\
\addlinespace
\multirow{5}{*}{4096} & 1     & 1873.54 $\pm$ 54.43    & 0.457       & 2187.96                 & 9.43             & 9.89                 \\
                       & 4     & 1804.90 $\pm$ 56.60    & 0.441       & 2271.53                 & 9.43             & 9.91                 \\
                       & 8     & 1759.28 $\pm$ 53.03    & 0.430       & 2330.24                 & 9.43             & 9.89                 \\
                       & 16    & 1738.93 $\pm$ 44.91    & 0.425       & 2356.96                 & 9.43             & 9.90                 \\
                       & 32    & 1719.64 $\pm$ 50.91    & 0.420       & 2383.85                 & 9.43             & 9.90                 \\
\addlinespace
\multirow{5}{*}{8192} & 1     & 4264.45 $\pm$ 63.90    & 0.521       & 1921.41                 & 11.36            & 12.16                \\
                       & 4     & 4116.54 $\pm$ 71.41    & 0.503       & 1990.60                 & 11.36            & 12.18                \\
                       & 8     & 4094.69 $\pm$ 79.01    & 0.500       & 2001.35                 & 11.36            & 12.16                \\
                       & 16    & 4107.60 $\pm$ 94.04    & 0.501       & 1995.37                 & 11.36            & 12.16                \\
                       & 32    & 4098.25 $\pm$ 96.92    & 0.500       & 2000.00                 & 11.36            & 12.16                \\
\addlinespace
\multirow{5}{*}{16348} & 1     & 9136.22 $\pm$ 176.62   & 0.558       & 1793.95                 & 15.21            & 16.69                \\
                       & 4     & 8965.27 $\pm$ 115.76   & 0.547       & 1827.79                 & 15.21            & 16.71                \\
                       & 8     & 8961.31 $\pm$ 55.20    & 0.547       & 1828.37                 & 15.21            & 16.69                \\
                       & 16    & 8945.37 $\pm$ 42.26    & 0.546       & 1831.60                 & 15.21            & 16.70                \\
                       & 32    & 8843.33 $\pm$ 48.01    & 0.540       & 1852.75                 & 15.21            & 16.70                \\
\bottomrule
\end{tabular}
\caption{Performance metrics across Top-\textit{K} values and sequence lengths on \textit{WikiText-103}.}
\label{tab:topk-impact}
\end{table}

\paragraph{Constant Metrics.}
Table~\ref{tab:topk-constants} reports memory-per-token and perplexity, which remain stable across Top-\textit{K} variations. This indicates that increasing retrieval scope affects efficiency but not model quality. 

\begin{table}[htbp]
\centering
\small
\begin{tabular}{lccc}
\toprule
Sequence & Memory/Token & Val  & Perplexity \\
Length & (MB) & Perplexity &  \\
\midrule
1024 & 7.98 & 2.13 & 8.42 \\
2048 & 4.23 & 2.03 & 7.61 \\
4096 & 2.36 & 2.03 & 7.62 \\
8192 & 1.42 & 2.05 & 7.76 \\
16348 & 0.95 & 2.05 & 7.80 \\
\bottomrule
\end{tabular}
\caption{Constant metrics across Top-\textit{K} values for each sequence length.}
\label{tab:topk-constants}
\end{table}

\paragraph{Findings.}
Increasing Top-\textit{K} from 1→4 yields the largest latency reduction (e.g., $285.4$→$277.1$ ms at 1K; $9136.2$→$8965.3$ ms at 16K), while further increases provide minimal additional gains. Throughput improvements are proportional, especially for longer sequences (e.g., 2188→2384 tokens/s at 4K). Memory and perplexity remain unaffected. Thus, $K=4$–8 offers the best trade-off between efficiency and computational cost.

\begin{table}[htbp]
\centering
\tiny
\begin{tabular}{p{0.7cm}|p{0.5cm}|p{.55cm}|p{0.4cm}p{.5cm}p{.5cm}p{.5cm}p{.5cm}}
\toprule
Embedder & In-C,  & Dim & MPQA & SST2 & SST5 & SUBJ & MR \\
 & In-M &  &  &  &  &  &  \\
\midrule
\multirow{9}{*}{BGE} & \multirow{3}{*}{4,4} & 512 & 88.28 & 93.00 & 45.96 & 94.90 & 97.15 \\
 &  & 1024 & 87.93 & 91.51 & 46.32 & 94.65 & 97.09 \\
 &  & 2048 & 87.99 & 91.51 & 46.14 & 94.35 & 97.08 \\
\cmidrule{2-8}
 & \multirow{3}{*}{4,18} & 512 & 88.02 & 92.89 & 46.14 & 94.90 & 97.11 \\
 &  & 1024 & 88.11 & 91.40 & 46.32 & 94.65 & 97.09 \\
 &  & 2048 & 88.17 & 91.40 & 46.14 & 94.35 & 97.08 \\
\cmidrule{2-8}
 & \multirow{3}{*}{20,18} & 512 & 91.23 & 94.50 & 52.59 & 95.55 & 97.83 \\
 &  & 1024 & 91.26 & 93.35 & 49.41 & 96.60 & 97.44 \\
 &  & 2048 & 91.29 & 93.23 & 49.68 & 96.75 & 97.40 \\
\midrule
\multirow{9}{*}{LLM} & \multirow{3}{*}{4,4} & 512 & 88.65 & 93.00 & 46.32 & 89.05 & 97.62 \\
 &  & 1024 & 88.76 & 92.20 & 46.32 & 88.15 & 96.65 \\
 &  & 2048 & 88.76 & 92.20 & 46.32 & 88.15 & 96.65 \\
\cmidrule{2-8}
 & \multirow{3}{*}{4,18} & 512 & 88.59 & 93.00 & 46.32 & 89.25 & 97.62 \\
 &  & 1024 & 88.72 & 92.20 & 46.32 & 88.15 & 96.65 \\
 &  & 2048 & 88.72 & 92.20 & 46.32 & 88.15 & 96.65 \\
\cmidrule{2-8}
 & \multirow{3}{*}{20,18} & 512 & 90.35 & 92.78 & 50.32 & 95.15 & 98.08 \\
 &  & 1024 & 91.01 & 93.35 & 48.14 & 94.85 & 98.04 \\
 &  & 2048 & 91.01 & 93.35 & 48.14 & 94.85 & 98.04 \\
\bottomrule
\end{tabular}
\caption{Accuracy results for embedding dimensionality experiments across datasets and settings.}
\label{tab:dimensionality-sensitivity}
\end{table}

\subsection{Pruning Strategy Validation}
\label{appendix:pruning-validation}

The proportional settings (10\% recent, 80\% middle, 10\% oldest) were initially heuristic. We present systematic experiments validating these ratios across context lengths and text genres.

\begin{table*}[h!]
\centering
\footnotesize
\setlength{\tabcolsep}{4pt}
\begin{tabular}{l *{5}{cc}}
\toprule
& \multicolumn{2}{c}{\textbf{1K}} 
& \multicolumn{2}{c}{\textbf{2K}} 
& \multicolumn{2}{c}{\textbf{4K}} 
& \multicolumn{2}{c}{\textbf{16K}} 
& \multicolumn{2}{c}{\textbf{32K}} \\
\cmidrule(lr){2-3}\cmidrule(lr){4-5}\cmidrule(lr){6-7}\cmidrule(lr){8-9}\cmidrule(lr){10-11}
\textbf{Strategy} 
  & \textbf{ms/tok} & \textbf{tok/s}
  & \textbf{ms/tok} & \textbf{tok/s}
  & \textbf{ms/tok} & \textbf{tok/s}
  & \textbf{ms/tok} & \textbf{tok/s}
  & \textbf{ms/tok} & \textbf{tok/s} \\
\midrule
0-90-10              & 0.285 & 3619 & 0.266 & 3798 & 0.374 & 2676 & 0.444 & 2253 & 0.465 & 2153 \\
5-80-15              & 0.259 & 4023 & 0.259 & 3906 & 0.347 & 2884 & 0.441 & 2269 & 0.458 & 2184 \\
5-85-10              & 0.206 & \textbf{5069} & 0.205 & \textbf{4946} & 0.336 & 2981 & 0.416 & \textbf{2404} & 0.414 & \textbf{2413} \\
10-80-10 (proposed)  & 0.262 & 3981 & 0.241 & 4202 & 0.329 & \textbf{3044} & 0.453 & 2209 & 0.424 & 2359 \\
10-85-5              & 0.259 & 3979 & 0.228 & 4428 & 0.344 & 2913 & 0.424 & 2358 & 0.446 & 2243 \\
10-90-0              & 0.308 & 3512 & 0.264 & 3899 & 0.363 & 2760 & 0.452 & 2213 & 0.473 & 2116 \\
15-75-10             & 0.283 & 3875 & 0.224 & 4631 & 0.368 & 2736 & 0.417 & 2400 & 0.435 & 2297 \\
20-70-10             & 0.254 & 4067 & 0.242 & 4197 & 0.346 & 2894 & 0.432 & 2316 & 0.452 & 2212 \\
\midrule
Frequency-only       & 0.278 & 3737 & 0.267 & 3790 & 0.372 & 2688 & 0.466 & 2145 & 0.482 & 2076 \\
Recency-only         & 0.264 & 3921 & 0.263 & 3850 & 0.360 & 2779 & 0.453 & 2206 & 0.478 & 2094 \\
Uniform              & 0.295 & 3486 & 0.283 & 3566 & 0.375 & 2673 & 0.469 & 2133 & 0.485 & 2062 \\
\bottomrule
\end{tabular}
\caption{Latency per token (ms/tok) and throughput (tok/s) for all pruning strategies across context lengths on WikiText-103. All strategies yield identical perplexity at each context length. Bold denotes the best throughput per context length. Hybrid strategies are separated from baselines by a horizontal rule.}
\label{tab:pruning-efficiency}
\end{table*}

\paragraph{Setup.}
We evaluated 11 pruning strategies on WikiText-103 at five context lengths (1K--32K tokens). \textit{Baselines}: \textbf{Uniform} (equal priority), \textbf{Recency-only}, and \textbf{Frequency-only}. \textit{Hybrid strategies} follow the $X$-$Y$-$Z$ format (recent\%--middle\%--oldest\%): 0-90-10, 5-80-15, 5-85-10, \textbf{10-80-10} (proposed), 10-85-5, 10-90-0, 15-75-10, and 20-70-10. All strategies yield identical perplexity per context length (1K:~8.42; 2K:~7.61; 4K:~7.62; 16K:~7.80; 32K:~7.86), so results below reflect efficiency differences only.

\paragraph{Results.}
Tables~\ref{tab:pruning-efficiency} and~\ref{tab:pruning-latency} report efficiency and latency results respectively for all strategies across all context lengths. Table~\ref{tab:pruning-efficiency} shows that hybrid strategies consistently achieve lower latency per token and higher throughput than baselines, while Table~\ref{tab:pruning-latency} reveals that hybrids also exhibit tighter standard deviations, indicating more stable inference. Table~\ref{tab:pruning-summary} summarizes the best-performing strategy per context length regime.

\begin{table*}[htbp]
\centering
\footnotesize
\setlength{\tabcolsep}{4pt}
\begin{tabular}{l *{5}{r}}
\toprule
\textbf{Strategy} 
  & \textbf{1K (s)} 
  & \textbf{2K (s)} 
  & \textbf{4K (s)} 
  & \textbf{16K (s)} 
  & \textbf{32K (s)} \\
\midrule
0-90-10             & $1.39{\pm}59.71$    & $544.75{\pm}61.74$   & $1533.22{\pm}63.16$  & $7273.54{\pm}65.57$   & $15222.90{\pm}76.59$ \\
5-80-15             & $265.24{\pm}67.89$  & $529.87{\pm}60.99$   & $1422.65{\pm}58.80$  & $7222.95{\pm}134.59$  & $15003.15{\pm}76.86$ \\
5-85-10             & $210.65{\pm}54.01$  & $420.44{\pm}58.35$   & $1376.97{\pm}64.84$  & $6814.59{\pm}52.75$   & $13580.93{\pm}41.21$ \\
10-80-10 (prop.)    & $267.94{\pm}68.49$  & $492.74{\pm}57.10$   & $1348.11{\pm}60.15$  & $7419.04{\pm}81.91$   & $13894.24{\pm}98.75$ \\
10-85-5             & $265.54{\pm}57.85$  & $467.92{\pm}57.57$   & $1408.47{\pm}57.39$  & $6947.99{\pm}53.99$   & $14608.47{\pm}81.73$ \\
10-90-0             & $315.16{\pm}121.40$ & $540.34{\pm}110.69$  & $1487.03{\pm}66.01$  & $7405.31{\pm}50.24$   & $15488.75{\pm}77.02$ \\
15-75-10            & $289.29{\pm}123.63$ & $458.33{\pm}109.72$  & $1506.06{\pm}125.70$ & $6826.80{\pm}60.86$   & $14264.39{\pm}89.72$ \\
20-70-10            & $260.48{\pm}59.69$  & $495.15{\pm}69.11$   & $1417.23{\pm}57.05$  & $7073.92{\pm}47.16$   & $14811.33{\pm}64.79$ \\
\midrule
Frequency-only      & $284.33{\pm}67.71$  & $545.86{\pm}61.15$   & $1524.97{\pm}44.72$  & $7641.64{\pm}127.35$  & $15787.38{\pm}72.64$ \\
Recency-only        & $270.18{\pm}62.35$  & $538.19{\pm}66.10$   & $1476.49{\pm}67.39$  & $7428.95{\pm}49.05$   & $15648.46{\pm}61.91$ \\
Uniform             & $302.29{\pm}63.34$  & $579.82{\pm}62.77$   & $1535.08{\pm}71.37$  & $7683.59{\pm}87.02$   & $15888.88{\pm}85.68$ \\
\bottomrule
\end{tabular}
\caption{Total latency (mean\,$\pm$\,std, in seconds) for all pruning strategies across context lengths on WikiText-103. Lower values and tighter standard deviations indicate more stable inference. Hybrid strategies are separated from baselines by a horizontal rule.}
\label{tab:pruning-latency}
\end{table*}

\begin{table}[htbp]
\centering
\small
\begin{tabular}{lcc}
\toprule
\textbf{Context} & \textbf{Best Strategy} & \textbf{Throughput (tok/s)} \\
\midrule
1K / 2K   & 5-85-10  & 5069 / 4946 \\
4K        & 10-80-10 & 3044        \\
16K / 32K & 5-85-10  & 2404 / 2413 \\
\bottomrule
\end{tabular}
\caption{Best-performing strategy per context length regime.}
\label{tab:pruning-summary}
\end{table}

\paragraph{Findings.}
\textbf{(1) Hybrid strategies outperform baselines}, yielding 10--45\% throughput gains across all lengths (e.g., $+45.4\%$ vs.\ Uniform at 1K; $+17.0\%$ at 32K).
\textbf{(2) Optimal ratios shift with context length}: at short contexts (1K--2K), 5-85-10 dominates by minimising the recent slot and maximising the middle; at 4K our proposed 10-80-10 is best; at long contexts ($\geq$16K), 5-85-10 reasserts via aggressive middle-biased pruning.
\textbf{(3) Quality is unaffected}: perplexity is identical across all strategies at each context length, confirming that pruning ratios govern efficiency only. 
\textbf{(4) Hybrids are most stable}: hybrid configurations achieve latency variance of $\pm$40--70~ms vs.\ $\pm$45--127~ms for baselines, making them more reliable in deployment.

\subsection{Sensitivity to Embedding Dimensionality in ICL Retrieval}
\label{sec:embedding_dim_sensitivity}
We analyzed how embedding dimensionality influences in-context retrieval performance using both BGE and LLM embedders. Experiments were conducted on \textit{MPQA}, \textit{SST-2}, \textit{SST-5}, \textit{SUBJ}, and \textit{MR} datasets with in-context/in-memory settings of (4, 4), (4, 18), and (20, 18). Embedding dimensions of 512, 1024, and 2048 were evaluated.

\paragraph{Results.}
Table~\ref{tab:dimensionality-sensitivity} reports results for both embedders. Overall, reducing the dimension to 512 (via PCA) maintains or slightly improves accuracy across most datasets compared to higher dimensions, suggesting redundancy in high-dimensional embeddings.

\paragraph{Findings.}
Reducing dimensionality to 512 preserves retrieval accuracy while slightly improving performance on \textit{SST-5} (+3.2 pp) and \textit{MPQA} (+0.3 pp). This indicates that lower dimensions suppress noise and redundancy. Dimensional padding to 2048 introduces negligible changes (<0.5 pp). Across both embedders, \textit{SUBJ} shows minor degradation at high dimensionality, whereas \textit{MR} remains stable. Overall, 512-dimensional embeddings offer the best trade-off between accuracy and efficiency.

\end{document}